\documentclass[preprint2]{emulateapj}

\newcommand{\reduceme}{\mbox{R\raisebox{-0.35ex}{E}D%
\hspace{-0.05em}\raisebox{0.85ex}{uc}\hspace{-0.90em}%
\raisebox{-.35ex}{{m}}\hspace{0.05em}E}}

\shorttitle{Kinematic Properties of dEs in Virgo}
\shortauthors{Toloba et al.}

\begin{document}


\title{Kinematic properties as probes of the evolution of dwarf galaxies in the Virgo cluster}


\author{E. Toloba\altaffilmark{1}}
\author{A. Boselli\altaffilmark{2}}
\author{J. Gorgas\altaffilmark{1}}
\author{R. F. Peletier\altaffilmark{3}}
\author{A. J. Cenarro\altaffilmark{4,8}}
\author{D. A. Gadotti\altaffilmark{5}}
\author{A. Gil de Paz\altaffilmark{1}}
\author{S. Pedraz\altaffilmark{6}}
\author{U. Yildiz\altaffilmark{3,7}}
\affil{$^1$Universidad Complutense de Madrid, 28040, Madrid, Spain}
\affil{$^2$Laboratoire d'Astrophysique de Marseille, UMR 6110 CNRS, 38 rue F. Joliot-Curie, 13388 Marseille, France}
\affil{$^3$Kapteyn Astronomical Institute, Postbus 800, 9700 AV Groningen, the Netherlands}
\affil{$^4$Instituto de Astrof\'{i}sica de Canarias, E-38200, La Laguna, Tenerife, Spain}
\affil{$^5$Max-Planck-Institut f\"{u}r Astrophysik, Karl-Schwarzschild-Str. 1, D-85748 Garching bei M\"{u}nchen, Germany}
\affil{$^6$Centro Astron\'{o}mico Hispano Alem\'{a}n, Calar Alto (CSIC-MPG), Almer\'{i}a, Spain}
\affil{$^7$Leiden Observatory, Leiden University, P.O. Box 9513, 2300 RA, Leiden, The Netherlands}
\affil{$^8$Centro de Estudios de F\'isica del Cosmos de Arag\'on, E-44001, Teruel, Spain}

\begin{abstract}

We present new observational results on the kinematical, morphological, and stellar population properties of a sample of 21 dEs located both in the Virgo cluster and in the field, which show that 52$\%$ of the dEs i) are rotationally supported, ii) exhibit structural signs of typical rotating systems such as discs, bars or spiral arms, iii) are younger ($\sim3$\,Gyr) than non-rotating dEs, and iv) are preferentially located either in the outskirts of Virgo or in the field.
This evidence is consistent with the idea that rotationally supported dwarfs are late type
spirals or irregulars that recently entered the cluster and lost their gas
through a ram pressure stripping event, quenching their star formation
and becoming dEs through passive evolution. We also find that all, but one, galaxies without photometric hints for hosting discs are pressure supported and 
 are all situated in the inner regions of the cluster. This suggests a different evolution from the rotationally supported systems. 
 Three different scenarios for these non-rotating galaxies are discussed (in situ formation, harassment and ram pressure stripping).

\end{abstract}

\keywords{galaxies: kinematics and dynamics --- galaxies: dwarf --- galaxies: clusters: individual (Virgo) --- galaxies: formation --- galaxies: evolution}

\section{Introduction}

Dwarf galaxies ($M_B>-18$)  are the most numerous objects in the  nearby Universe. They have gained importance since hierarchical models proposed dwarfs as the building blocks of massive galaxies \citep[e.g.,][]{WR78,WF91}. These systems can be divided in star forming (Im, BCD, Sd,...) and quiescent (dE, dS0) dwarfs, the latter being the dominant population in clusters, the former  the most common in the field \citep{Sand85,FB94,Blant05,Crot05}. The study of the population of low-mass galaxies and of the mechanisms leading to the strong morphology segregation between the different types of dwarfs is fundamental to understand the assembly and evolution of the overall population of galaxies.

In the case of dEs, various scenarios have been suggested to explain their formation.
Whether they are the low luminosity extension of the giant ellipticals (Es) or they are the result of different formation and evolution processes is still an open question \citep{FB94}. It has been proposed that dEs were late-type systems where the interstellar medium (ISM)  was swept away by the kinetic pressure due to supernova explosions (\citet{YosAri87}; see however \citet{ST01}).
Other theories suggested that late-type spirals stopped their star formation once their ISM was removed during their interaction with the environment and evolved into quiescent dwarfs. The existence of a morphological segregation effect on the dwarf galaxy population \citep{FB94} favors this second scenario. Different processes could be at the origin of gas removal in clusters: interaction with the intergalactic medium (IGM) \citep[e.g.,][]{VZ04}, as ram-pressure stripping, galaxy-galaxy interactions \citep[e.g.,][]{ByrdValt90} and harassment \citep[e.g.,][]{Moore98, Mast05}. These mechanisms are  able to reproduce some of the observed properties of local dEs in clusters. Structural parameters, such as surface brightness, and spectrophotometric properties, such as stellar populations, are easily reproduced after a ram pressure event \citep{Boselli08a,Boselli08b}. The detailed study of \citep[][]{Lisk06b,Lisk06a,Lisk07} shows that there exists a population of galaxies that have properties in between those of dEs and star forming dwarfs. In these objects, which are classified as dEs, some remains of spiral discs, such as spiral arms or irregular features, are still visible. If dEs are formed from star forming galaxies through ram pressure stripping, we expect that the angular momentum of the parent galaxies should be conserved, while in the case of multiple dynamical perturbations the system is rapidly heated and the rotation is lost. Measuring the kinematics of dEs is therefore an important test to understand their origin.

In the last decade considerable efforts have been made to measure the kinematic properties of dE in Virgo, the closest rich cluster
\citep[e.g.,][]{Ped02,Geha02,Geha03,VZ04}. These works have shown the existence of both rotating and pressure supported systems. This fact, together with the varieties of morphologies found by \citet{Lisk06b} makes it clear that the formation of dEs might rather be complex. Would every different population of dEs have a separate formation process?
To answer this question we started a kinematical survey of dE in the Virgo cluster.This is the first of a series of papers devoted to the study of the kinematics and stellar populations of dEs in the Virgo cluster.

\section{The Data}

\subsection{Observations and Data Reduction}

The data were obtained as part of the MAGPOP-ITP collaboration (Multiwavelength Analysis of Galaxy Populations-International Time Program), a  Marie Curie Research Training Network. 
The observed sample,  selected from the Sloan Digital Sky Survey  (SDSS) as described in \citet{Mich08}, contains 18 Virgo and 3 field dEs. The Virgo galaxies were chosen to have $m_B>15$ and dE or dS0 classification in the Virgo Cluster Catalog  \citet[VCC,][]{Bing85}, with available GALEX data \citep{Boselli05}. The field sample was required to be in the SDSS velocity range 375km$~$s$^{-1}$$<v_{hel}<$1875km$~$s$^{-1}$ and $-$18.5$<M_r'<-$15 mag. Quiescent galaxies were selected to have  FUV-NUV$>$0.9 or $u-g>$1.2 when there were no UV detections. 

We obtained medium resolution ($R\simeq3800$) long-slit spectroscopy 
along the major axis of 21 dEs  during three different runs at 
Roque de los Muchachos
Observatory, Canary Islands: two at the WHT (4.2m), using the spectrograph ISIS
($3445-8950$\AA), and one at the INT (2.5m) with the IDS ($4600-5960$\AA). 
With a slit width of 2$"$and exposure times of 1 hour/configuration, the gratings used in each campaign were R1200B for IDS and R1200B and R600R for ISIS. ISIS has the possibility to observe with a dichroic and split the light into two beams to cover a larger wavelength range. Only in the first run we also used the mirror to cover the range $4800-5600$\AA. The spectral resolution obtained was 1.6\AA(FWHM) (49km$~$s$^{-1}$) and 3.2\AA(FWHM) (58km$~$s$^{-1}$)  in the blue and red arms of ISIS respectively, and 1.8\AA(FWHM) (45km$~$s$^{-1}$) with IDS.

The reduction was done using standard procedures for long-slit spectra in the optical range, using \reduceme \citep[][]{Car99}, a package specially focused on the parallel treatment of errors. The flux calibration was done using the stars observed from the MILES and CaT libraries \citep[][respectively]{SB06lib,cen01}. More details of the observations will be presented in Toloba et al. in preparation (Paper I).

\subsection{Kinematic and photometric parameters}

To calculate the stellar kinematics we used MOVEL, available in \reduceme.
MOVEL is an algorithm based on the Fourier quotient method described by \citet{Sarg77} and improved with OPTEMA \citep[][]{Gon93}, that allows us to overcome the typical template mismatch problem. MOVEL uses an iterative procedure to determine the radial velocity and the stellar broadening of the galaxy, by fitting a galaxy model, created as a linear combination of the stars introduced as templates, to the data.

To make sure that the stars used as templates have the same instrumental profile as the target galaxies, they were defocused to fill the slit in the same way as the galaxies. In Paper I more details are given about how this was exactly done.

\begin{figure*}
\centering
\resizebox{0.45\textwidth}{!}{\includegraphics[angle=-90]{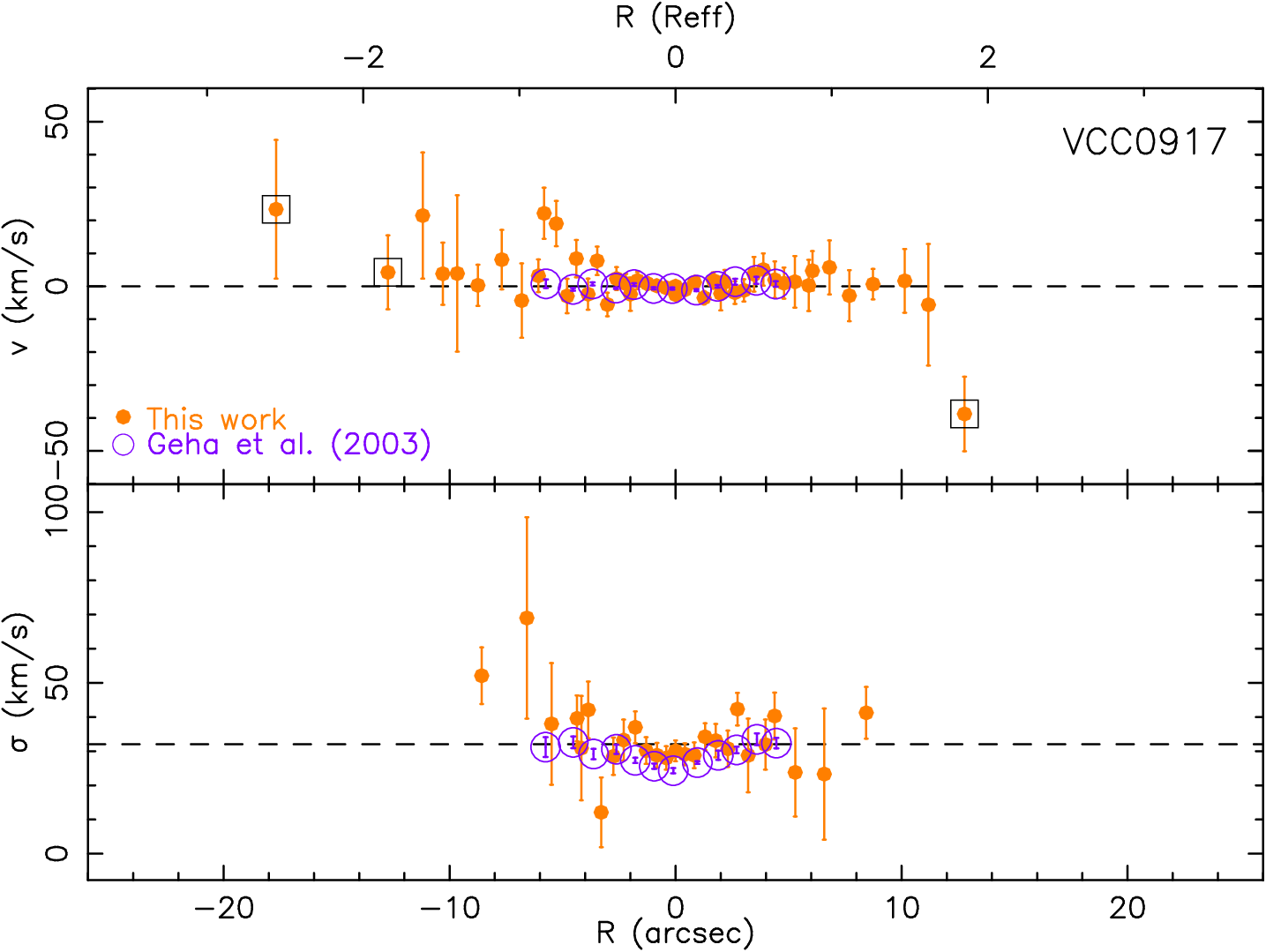}}
\resizebox{0.45\textwidth}{!}{\includegraphics[angle=-90]{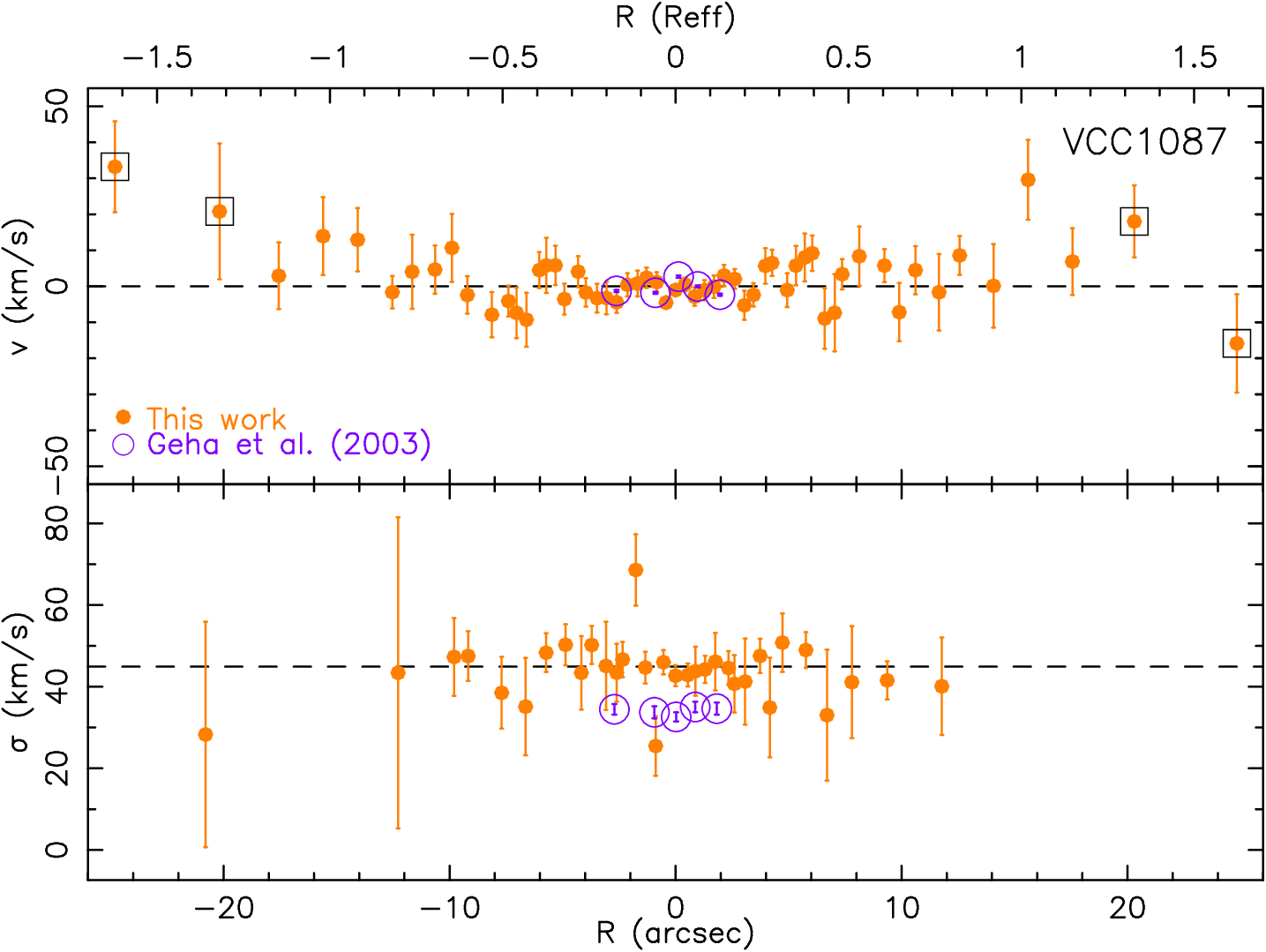}}
\caption{Example of the kinematical profiles obtained. The open squares show the values used to calculate  $v_{max}$ and the dashed line the lower pannel, the central $\sigma$.
\label{f0}}
\end{figure*}

The maximum rotational velocity ($v_{max}$) was calculated as the weighted-average of the 2 highest velocities along the major axis on both sides of the galaxy at the same radius (for non-symmetric profiles at least 3 values are required), typically located at around 10$"$ from the center (Figure \ref{f0}). After de-redshifting the spectra using the rotation curves, the central velocity dispersion ($\sigma$) has been computed coadding all the individual spectra up to 1 effective radius. The typical $S/N$ for the central $\sigma$ is $\sim$60$/$\AA,  while $S/N\ge$10 for $v_{max}$ at 10$"$.

The photometric parameters: the effective radius ($R_{eff}$), the
ellipticity ($\epsilon$) at $R_{eff}$ and the boxyness/diskyness parameter
($A_4/A$), measured in a radius range from 5$"$  to $4R_{eff}$, were calculated using the IRAF task {\sc ellipse}
over the H and K band photometry observed within the MAGPOP
collaboration (Peletier et al. in preparation).

\section{Results}

Given the radial decrease of the galaxy surface brightness, the $S/N$ in the
outer parts of some of the galaxies was not enough to reach a plateau of the
rotation curves. To be cautious, for those galaxies where the $v_{max}$  is measured at a radius $<6''$, the value is considered as a lower limit.

\begin{figure*}
\centering
\resizebox{0.7\textwidth}{!}{\includegraphics[angle=-90]{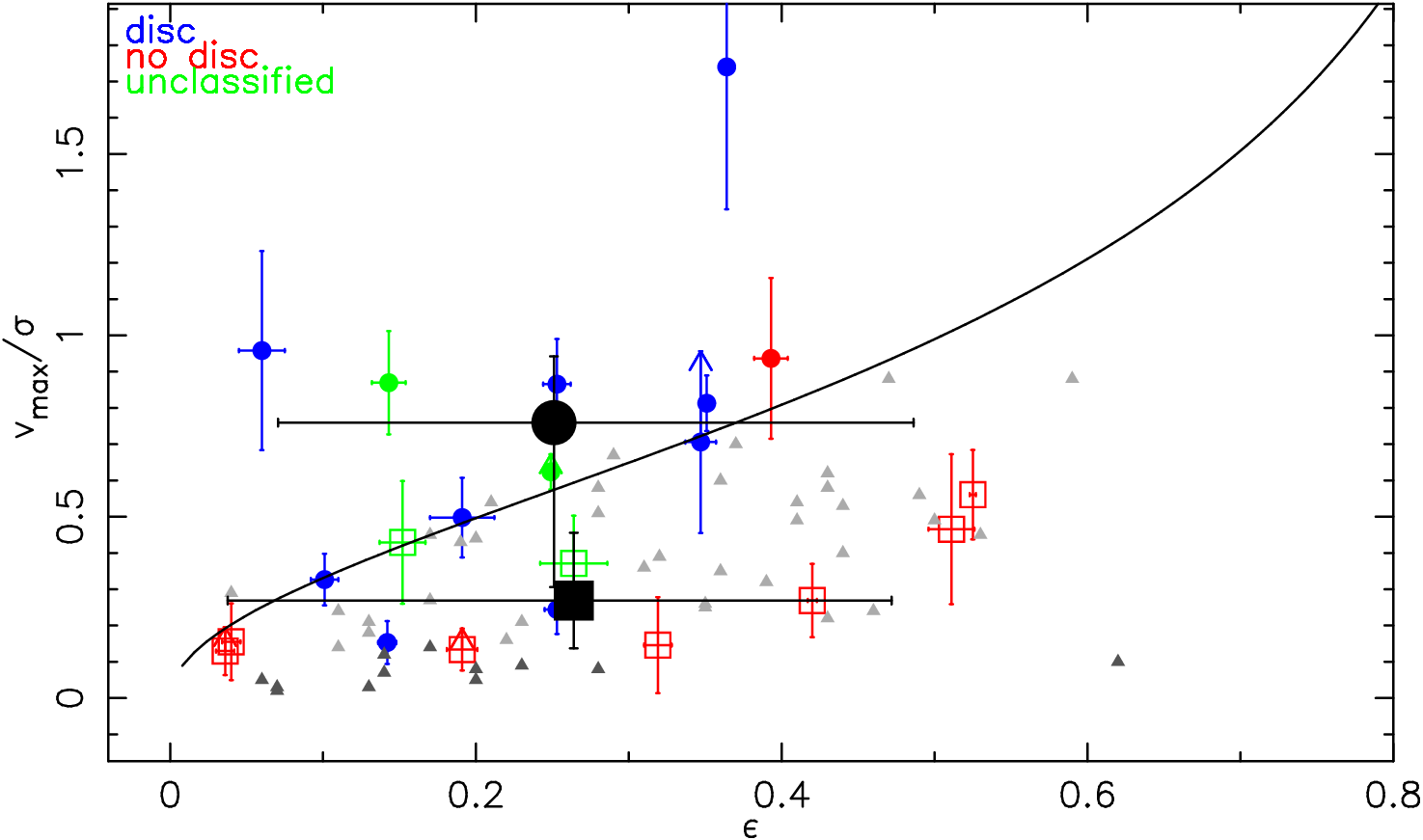}}
\caption{Anisotropy diagram. The solid line is the model for an
isotropic oblate system flattened by rotation \citep{Binn78}. Triangles are 
giant ellipticals (from \citet{Em07}), with slow rotators ($v_{max}/\sigma<0.1$) in dark grey, and fast rotators ($v_{max}/\sigma>0.1$) in light grey. 
Blue, red and green symbols show our sample of dEs as classified by \citet{Lisk06b}. The
filled dots and the open squares indicate
galaxies with and without discs on the basis of 
$A_4/A$. Lower limits on $v_{max}$ are indicated with arrows. The black symbols represent the median 
for dEs with (black dot)/without disc (black square) based on 
$A_4/A$ classification, their error bars contain 34.1$\%$ ($\sim1\sigma$) of the values at both sides of the median. 
\label{f1}}
\end{figure*}

Figure \ref{f1} shows the anisotropy diagram. $v_{max}/\sigma$ is the ratio between the maximum rotational velocity and the central velocity dispersion of the galaxies. The colour code corresponds to the classification of
\citet{Lisk06b}. In blue are all the dEs that have signs
indicating a possible spiral origin, those classified as certain, probable and
possible discs (spiral arms, edge-on discs or bars) and also those classified
as ''other'' (mainly irregular central features).  In red are
indicated those that do not
show any underlying structure and in green the galaxies that were not
in the sample of \citet{Lisk06b}; those are the 3 field dEs and VCC1947.  We do not distinguish between galaxies with and without blue centers
\citep{Lisk06a} and between nucleated and non-nucleated dwarfs
\citep{Bing85} due to the lack of statistics, with only 3 galaxies with a blue
nucleus and 3 non$-$nucleated objects.

The average $A_4/A$ parameter, which measures how different the isophotes are from
a perfect ellipse ($A_4/A>0$ means disky isophotes, $A_4/A<0$  boxy isophotes \citep{KorBen96}),
shows that all disky galaxies (except two) have been classified by
\citet{Lisk06b} as discs. The two exceptions are (see Table \ref{t1}):
VCC0917 where \citet{Lisk06b} did not see the disc revealed by the disky isophotes,
and VCC0308 where the low ellipticity suggests that
the galaxy is face-on, so $A_4/A$ can not accurately determine the underlying
features (the unsharp masking technique of  \citet{Lisk06b}
is here more useful). Regarding the four galaxies not included in
the sample of \citet{Lisk06b} (Table \ref{t1}) $A_4/A$ suggests that the
two with higher $v_{max}/\sigma$ contain discs (PGC1007217 and  VCC1947), while
the two with lower $v_{max}/\sigma$ do not (NGC3073 and  PGC1154903). 

\begin{table}
\begin{center}
\caption{$A_4/A$ Alternative classification. \label{t1}}
\begin{tabular}{|c|c|c|c|c|c|}
\tableline 
Galaxy      &   $v_{max}/\sigma$          & $\epsilon$      &   100x$A_4/A$       & $A_4/A$ \\
                &                                        &                         &                                & Classification \\
\tableline \tableline
PGC1007217  &   0.87$\pm$0.14 &   0.14$\pm$0.01  &   0.48$\pm$0.11    & disc               \\
PGC1154903\footnote{In NED: 2MASXJ02420036+0000531}     & 0.37$\pm$0.13    &   0.26$\pm$0.02     &   0.02$\pm$0.35    & no disc          \\
NGC3073 &  0.43$\pm$0.17    &   0.15$\pm$0.02     &   -0.35$\pm$0.03   & no disc          \\
VCC0308  &   0.96$\pm$0.27   &    0.06$\pm$0.02     &  -0.09$\pm$0.07   & no disc          \\
VCC0917  &   0.94$\pm$0.22   &    0.39$\pm$0.01     &   6.32$\pm$1.63    & disc                \\
VCC1947  &   0.62$\pm$0.05   &    0.25$\pm$0.01     &   0.35$\pm$0.05    & disc                \\
\tableline
\end{tabular}
\begin{flushleft}
Note: Galaxies without morphological classification in \citet{Lisk06b} and VCC0308 $\&$ VCC0917 whose classification by $A_4/A$ disagrees with \citet{Lisk06b}.
\end{flushleft}
\end{center}
\end{table}

The anisotropy diagram (Figure \ref{f1}) separates rotationally supported galaxies (above the solid line) from pressure dominated (below the solid line). Figure \ref{f1} shows that: 1) as ellipticals, dEs can be separated into slow ($v_{max}/\sigma<$0.1) and fast ($v_{max}/\sigma>$0.1) rotators
\citep{Em07}; 2) a large fraction (52$\%$) of the dEs are rotationally supported and 3) all of the rotationally supported galaxies have morphological and/or photometric signs of a spiral origin.

The apparent discordance with 
\citet{Geha03}, who found that the majority of the dEs were not rotating, is due to the fact that their data are limited to the core of the galaxies, never reaching radii larger than 6$''$ (Figure \ref{f0}), where the increase of the rotational velocity is generally observed (Paper I). A maximum offset of 10km$~$s$^{-1}$ in $\sigma$ with respect to \citet{Geha03} is found; however, it does not affect this result, since it makes $v_{max}/\sigma$ larger. This discrepancy could be due to the use of a K-type star template in the case of \citet{Geha03}, whereas a linear combination of different spectral types (from B to M in our case) is more appropriate for dEs.  Also a wider slit (2'' compared to 0.5'' in \citet{Geha03}) could increase the $\sigma$ measured if there were some ordered motion along the minor axis.

\begin{figure*}
\centering
\resizebox{0.7\textwidth}{!}{\includegraphics[angle=-90]{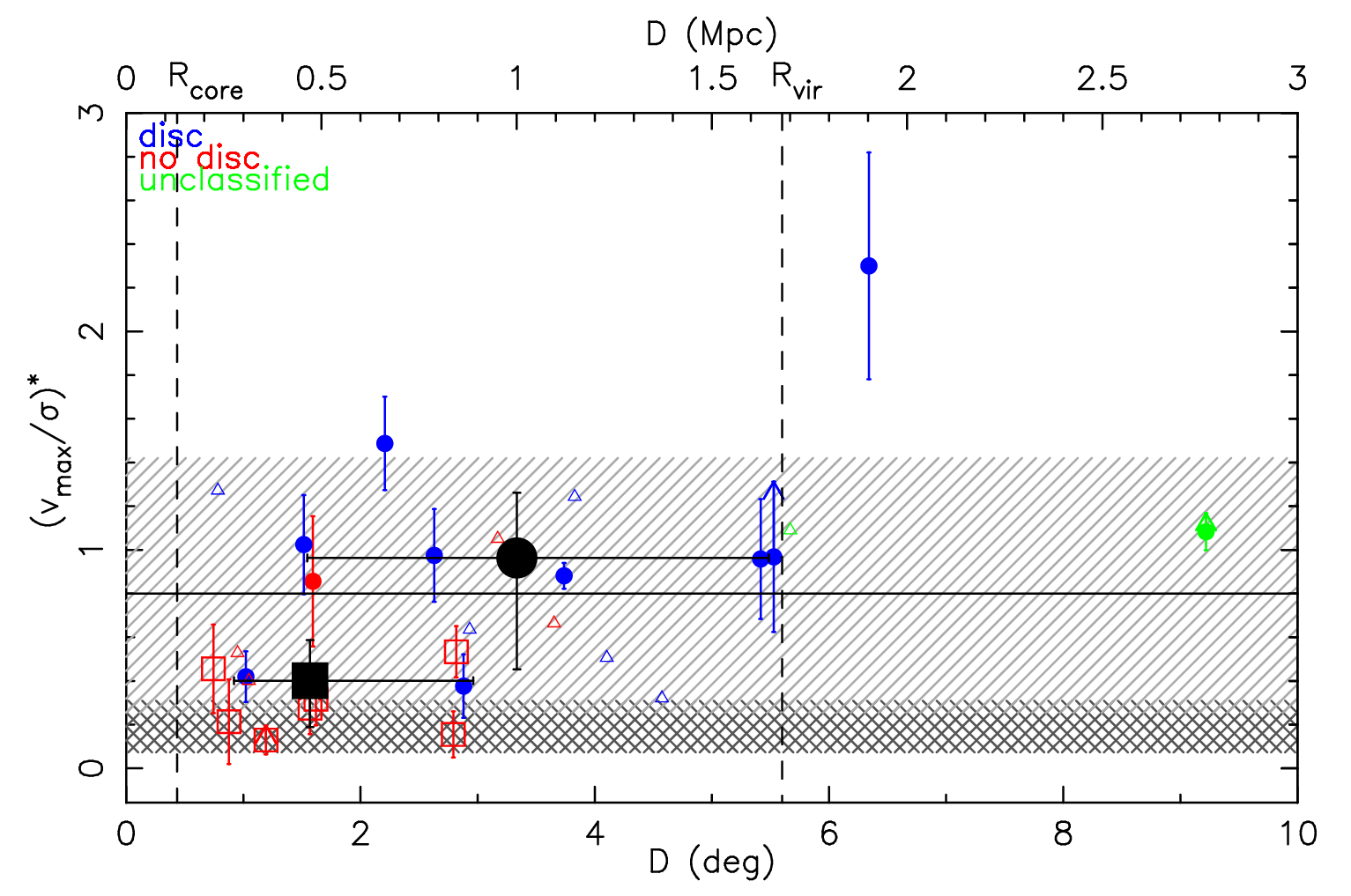}}
\caption{Anisotropy parameter versus the angular distance to M87 (centre of
Virgo). The symbols are as in Figure \ref{f1}. The light dark grey rectangles limit the $(v_{max}/\sigma)^*$ regions for fast and slow rotating Es as defined by \citet{Em07}. The open triangles
are \citet{VZ04}, included in the median values. The dashed lines show the Virgo core radius (130 kpc) and
virial radius (1.68Mpc) \citep{BG06}. The 
solid line divides the diagram into rotationally   
($(v_{max}/\sigma)^*>0.8$) and pressure supported galaxies ($(v_{max}/\sigma)^*<0.8$).\label{f2}}
\end{figure*}

Figure \ref{f2} shows $(v_{max}/\sigma)^*=\frac{v_{max}/\sigma}{\sqrt{\epsilon/(1-\epsilon)}}$ as a function of the projected angular distance from the cluster center (we assume Virgo to be at 17Mpc, \citet[][]{GB99}). $(v_{max}/\sigma)^*$ is $v_{max}/\sigma$ divided by the isotropic oblate model, which means that for values of $(v_{max}/\sigma)^*>0.8$ the galaxies are rotationally supported.  To improve statistics we have included the dEs of \citet{VZ04} (open triangles), whose data are consistent with ours (see Paper I). 

Rotationally supported systems are generally located in the cluster outskirts or in the field, while the $\sigma$-dominated dwarfs are found only in the central regions of the cluster. This idea had been suggested before
\citep{Geha03,VZ04}, but no clear confirmation was found. \citet{VZ04}
studied the inner 6 degrees (radius) of Virgo, 
and found a hint that very slow
rotators or non-rotating dEs were located  in the core or in the highest density
clumps. Our observations reach distances further away from M87, and the evidence of a trend is clearer. The non-parametric Spearman statistical test finds a correlation ($r=$0.54) with a confidence of 98.9$\%$.
Indeed a significant difference in the two dwarf galaxy populations can be seen
also in their median values, with dE with signs of a spiral origin having systematically
higher velocity rotations and being situated at the periphery of the cluster
($(v_{max}/\sigma)^*=0.96^{+0.30}_{-0.51}$, $D=0.99^{+0.64}_{-0.53}$ Mpc) with respect to non-disc dE ($(v_{max}/\sigma)^*=0.40^{+0.19}_{-0.21}$, $D=0.47^{+0.41}_{-0.19}$ Mpc). We emphasize that
any relation with the cluster-centric distance is smeared out by projection effects.

\begin{figure*}
\centering
\resizebox{0.7\textwidth}{!}{\includegraphics[angle=-90]{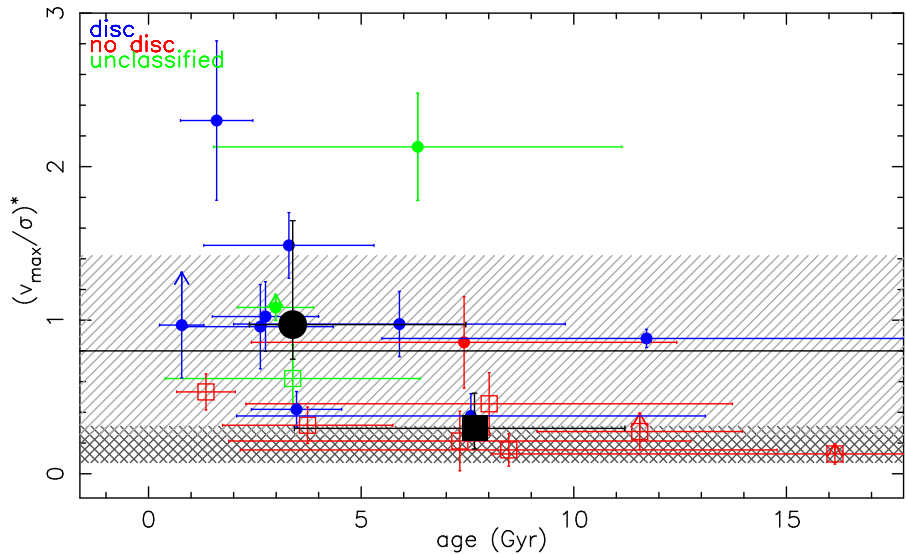}}
\caption{Anisotropy parameter versus the luminosity-weighted age from \citet{Mich08}. Symbols and shaded areas are as in Figures \ref{f1} and \ref{f2}.\label{f3}}
\end{figure*}

Figure \ref{f3} shows the anisotropy parameter as a function
of the age of the galaxies from \citet{Mich08}. The Spearman non-parametric test
finds a correlation ($r=-$0.64) with a 99.9$\%$ of confidence. Indeed the median values for the two populations are significantly
different; the dEs with discs are $\sim$3Gyr younger than the others. 
Figure \ref{f3} also shows that while dEs with no signs of a spiral origin 
might be of all ages but on average old (5$/$8 of the squares are older than 7 Gyr), disc galaxies are preferentially young (9$/$12 of the dots are younger than 7 Gyr), suggesting that the 2 populations might have a different origin.

To conclude, our results suggest that the rotationally supported dEs are preferentially 
located in the field or in the outskirts of the cluster and are on average
younger than those that are pressure supported, with no signs of spiral features, generally situated in the core of the cluster.

\section{Discussion}

It is believed that the difference between fast and slow rotators in Es is due to a different formation origin, with slow rotators, the most massive objects, principally located in dense environments and resulting from major mergers, while fast rotators, being less massive and more gas$-$rich \citep{Em07}, have a star formation history that is more extended in time. For dwarf galaxies this segregation in mass is difficult to be appreciated because of their small dynamical range (8.8$<L_H/L_{H\odot}<$9.7 in our sample). Moreover, models indicate that the probability that dEs have undergone a major merging event is almost null \citep{deLucia06}. This is consistent with the fact that dEs are on average much younger than Es \citep{Mich08}, whose formation happened probably at $z>$2 \citep{Renz06}. If their star formation activity was quenched in recent epochs, this would hardly be due to major merging events since the high $\sigma$ of clusters already formed prevented strong dynamical interactions \citep{BG06}.

Dwarf elliptical systems might have a different origin according to their structural, spectrophotometric and kinematical properties. Although harassment would be able to explain the origin of the pressure supported dwarfs, it is not likely to explain all the properties of rotationally supported systems ($(v_{max}/\sigma)^*>$0.8). Harassment has long time scales since it needs multiple encounters to be efficient in removing gas and stars and quenching the star formation \citep{BG06}, a scenario inconsistent with the observations which indicate that the star formation was active down to recent epochs, as the luminosity-weighted ages derived by \citet{Mich08} suggest (see Figure \ref{f3}). Furthermore, dynamical interactions heat the system removing rotation and enhancing velocity dispersion. In the simulations of \citet{Mast05} those galaxies that keep rotation after the harassment still retain an important spiral structure, yet they can not be classified as ellipticals. Moreover harassment is more efficient in the core of the clusters (in the inner 100kpc), while our result indicates that these rotationally supported systems are located mostly in the outskirts.
All these evidences are however consistent with a recent ram pressure stripping event as proposed by \citet{Boselli08a,Boselli08b}. Indeed, ram pressure
would easily remove the total gas content of the infalling low luminosity,
late-type galaxies, quenching their star formation, but conserving, at least on
short time scales, their angular momentum and therefore their rotation. The spectrophotometric and structural
properties of dE in Virgo are consistent with a recent ($\le$2 Gyr) interaction
with  the cluster hot IGM \citep{Boselli08a,Boselli08b}. Consistent with observations \citep{CK08}, hydrodynamical
simulations indicate that ram pressure stripping can be efficient up to the
virial radius, or even more outside in dwarf systems with shallow potential wells as
those analysed here \citep{Roed08}.

Pressure supported galaxies ($(v_{max}/\sigma)^*<0.8$) might have a different origin:  1) they might be galaxies formed in situ through the isotropic collapse of the gas at early epochs, thus not transformed by the environment, and be the extension of Es, with the exception that they probably did not undergo major merging events \citep{deLucia06}. Indeed they have old stellar populations, they do not present spiral structures \citep{Lisk06b} and they are virialised within the cluster, thus are members since early epochs ~\citep{Conselice01}. 2) They can be star forming systems that entered into the cluster in the early epochs, maybe through the accretion of groups where preprocessing was active, and later modified by galaxy harassment or 3) they are star forming systems that entered into the cluster several Gyr ago, where ram pressure, although less efficient than today, because of the lower density of the IGM and of the smaller velocity dispersion of the cluster still in formation, had the time through multiple cluster crossing to remove the gas and stop the star formation activity.  The lack of supply of new generations of stars on the plane of the disc increases $\sigma$  on long time scales, while a decrease of the rotation would in any case require dynamical interactions. The 2 $\sigma$-dominated disc dwarfs (VCC1183, VCC1910) could be galaxies in the last stages of their transformation into dEs.

The analysis done so far, although crucial for understanding the origin of the rotationally supported systems, is still insufficient to explain the origin of the $\sigma$ supported galaxies. It is probable that harassment and ram pressure  were acting together with a relative weight that might have changed
with time. Observations of high redshift clusters, mainly populated by massive, red sequence galaxies at late epochs \citep{deLucia06} consistently suggest that harassment and ram pressure are the most probable effects. It is however clear that a complete study of the kinematical, structural and spectrophotometric properties of early and late type galaxies in clusters combined with model predictions, is a powerful way to study the role of the environment on their evolution, a work we plan to do in the near future.

In summary, we report the first clear evidence of a correlation between the presence of rotationally supported dEs and the distance to the center of the Virgo cluster. We have seen that the further away a dwarf galaxy is from M87, the larger its rotation is and the younger it appears. In the outer parts many of them show spiral features beside their elliptical appearance. These kinematical data are consistent with the idea of ram pressure stripping transforming dwarf star forming galaxies into dEs.
Dwarf ellipticals that are pressure supported and show no sign of spiral features are likely to be
late-type systems that entered in the cluster at early epochs and were later transformed
by a combined effect of multiple encounters with nearby companions and
the interaction with the hot and dense IGM. They might also be originally
formed in clusters from an isotropic collapse of the primordial gas cloud.

\acknowledgments

We thank the MAGPOP EU Marie Curie Training and Research Network for financial support and the anonymous referee for useful comments. ET and AGdP thank the Spanish research projects AYA2007-67752-C03-03 and AYA2006-02358 respectively. A.J.C. is Juan de la Cierva Fellow of the Spanish Ministry of Science and Innovation.
This paper made use of the following public databases: SDSS, NED, HyperLEDA, GOLDMine \citep{GB03}.

\bibliographystyle{aa}
\bibliography{references}{}

\begin{thebibliography}{39}
\expandafter\ifx\csname natexlab\endcsname\relax\def\natexlab#1{#1}\fi

\bibitem[{{Binggeli} {et~al.}(1985){Binggeli}, {Sandage}, \&
  {Tammann}}]{Bing85}
{Binggeli}, B., {Sandage}, A., \& {Tammann}, G.~A. 1985, \aj, 90, 1681

\bibitem[{{Binney}(1978)}]{Binn78}
{Binney}, J. 1978, \mnras, 183, 501

\bibitem[{{Blanton} {et~al.}(2005){Blanton}, {Lupton}, {Schlegel}, {Strauss},
  {Brinkmann}, {Fukugita}, \& {Loveday}}]{Blant05}
{Blanton}, M.~R., {Lupton}, R.~H., {Schlegel}, D.~J., {et~al.} 2005, \apj, 631,
  208

\bibitem[{{Boselli} {et~al.}(2008{\natexlab{a}}){Boselli}, {Boissier},
  {Cortese}, \& {Gavazzi}}]{Boselli08a}
{Boselli}, A., {Boissier}, S., {Cortese}, L., \& {Gavazzi}, G.
  2008{\natexlab{a}}, \apj, 674, 742

\bibitem[{{Boselli} {et~al.}(2008{\natexlab{b}}){Boselli}, {Boissier},
  {Cortese}, \& {Gavazzi}}]{Boselli08b}
---. 2008{\natexlab{b}}, \aap, 489, 1015

\bibitem[{{Boselli} {et~al.}(2005){Boselli}, {Cortese}, {Deharveng}, {Gavazzi},
  {Yi}, {Gil de Paz}, {Seibert}, {Boissier}, {Donas}, {Lee}, {Madore},
  {Martin}, {Rich}, \& {Sohn}}]{Boselli05}
{Boselli}, A., {Cortese}, L., {Deharveng}, J.~M., {et~al.} 2005, \apjl, 629,
  L29

\bibitem[{{Boselli} \& {Gavazzi}(2006)}]{BG06}
{Boselli}, A. \& {Gavazzi}, G. 2006, \pasp, 118, 517

\bibitem[{{Byrd} \& {Valtonen}(1990)}]{ByrdValt90}
{Byrd}, G. \& {Valtonen}, M. 1990, \apj, 350, 89

\bibitem[{{Cardiel}(1999)}]{Car99}
{Cardiel}, N. 1999, PhD thesis, Universidad Complutense de Madrid, Spain

\bibitem[{{Cenarro} {et~al.}(2001){Cenarro}, {Cardiel}, {Gorgas}, {Peletier},
  {Vazdekis}, \& {Prada}}]{cen01}
{Cenarro}, A.~J., {Cardiel}, N., {Gorgas}, J., {et~al.} 2001, \mnras, 326, 959

\bibitem[{{Conselice} {et~al.}(2001){Conselice}, {Gallagher}, \&
  {Wyse}}]{Conselice01}
{Conselice}, C.~J., {Gallagher}, III, J.~S., \& {Wyse}, R.~F.~G. 2001, \apj,
  559, 791

\bibitem[{{Croton} {et~al.}(2005){Croton}, {Farrar}, {Norberg}, {Colless},
  {Peacock}, {Baldry}, {Baugh}, {Bland-Hawthorn}, {Bridges}, {Cannon}, {Cole},
  {Collins}, {Couch}, {Dalton}, {De Propris}, {Driver}, {Efstathiou}, {Ellis},
  {Frenk}, {Glazebrook}, {Jackson}, {Lahav}, {Lewis}, {Lumsden}, {Maddox},
  {Madgwick}, {Peterson}, {Sutherland}, \& {Taylor}}]{Crot05}
{Croton}, D.~J., {Farrar}, G.~R., {Norberg}, P., {et~al.} 2005, \mnras, 356,
  1155

\bibitem[{{Crowl} \& {Kenney}(2008)}]{CK08}
{Crowl}, H.~H. \& {Kenney}, J.~D.~P. 2008, \aj, 136, 1623

\bibitem[{{De Lucia} {et~al.}(2006){De Lucia}, {Springel}, {White}, {Croton},
  \& {Kauffmann}}]{deLucia06}
{De Lucia}, G., {Springel}, V., {White}, S.~D.~M., {Croton}, D., \&
  {Kauffmann}, G. 2006, \mnras, 366, 499

\bibitem[{{Emsellem} {et~al.}(2007){Emsellem}, {Cappellari}, {Krajnovi{\'c}},
  {van de Ven}, {Bacon}, {Bureau}, {Davies}, {de Zeeuw}, {Falc{\'o}n-Barroso},
  {Kuntschner}, {McDermid}, {Peletier}, \& {Sarzi}}]{Em07}
{Emsellem}, E., {Cappellari}, M., {Krajnovi{\'c}}, D., {et~al.} 2007, \mnras,
  379, 401

\bibitem[{{Ferguson} \& {Binggeli}(1994)}]{FB94}
{Ferguson}, H.~C. \& {Binggeli}, B. 1994, \aapr, 6, 67

\bibitem[{{Gavazzi} {et~al.}(2003){Gavazzi}, {Boselli}, {Donati}, {Franzetti},
  \& {Scodeggio}}]{GB03}
{Gavazzi}, G., {Boselli}, A., {Donati}, A., {Franzetti}, P., \& {Scodeggio}, M.
  2003, \aap, 400, 451

\bibitem[{{Gavazzi} {et~al.}(1999){Gavazzi}, {Boselli}, {Scodeggio}, {Pierini},
  \& {Belsole}}]{GB99}
{Gavazzi}, G., {Boselli}, A., {Scodeggio}, M., {Pierini}, D., \& {Belsole}, E.
  1999, \mnras, 304, 595

\bibitem[{{Geha} {et~al.}(2002){Geha}, {Guhathakurta}, \& {van der
  Marel}}]{Geha02}
{Geha}, M., {Guhathakurta}, P., \& {van der Marel}, R.~P. 2002, \aj, 124, 3073

\bibitem[{{Geha} {et~al.}(2003){Geha}, {Guhathakurta}, \& {van der
  Marel}}]{Geha03}
---. 2003, \aj, 126, 1794

\bibitem[{{Gonz{\'a}lez}(1993)}]{Gon93}
{Gonz{\'a}lez}, J.~J. 1993, PhD thesis, Thesis (PH.D.)--UNIVERSITY OF
  CALIFORNIA, SANTA CRUZ, 1993.Source: Dissertation Abstracts International,
  Volume: 54-05, Section: B, page: 2551.

\bibitem[{{Kormendy} \& {Bender}(1996)}]{KorBen96}
{Kormendy}, J. \& {Bender}, R. 1996, \apjl, 464, L119+

\bibitem[{{Lisker} {et~al.}(2006{\natexlab{a}}){Lisker}, {Glatt}, {Westera}, \&
  {Grebel}}]{Lisk06b}
{Lisker}, T., {Glatt}, K., {Westera}, P., \& {Grebel}, E.~K.
  2006{\natexlab{a}}, \aj, 132, 2432

\bibitem[{{Lisker} {et~al.}(2006{\natexlab{b}}){Lisker}, {Grebel}, \&
  {Binggeli}}]{Lisk06a}
{Lisker}, T., {Grebel}, E.~K., \& {Binggeli}, B. 2006{\natexlab{b}}, \aj, 132,
  497

\bibitem[{{Lisker} {et~al.}(2007){Lisker}, {Grebel}, {Binggeli}, \&
  {Glatt}}]{Lisk07}
{Lisker}, T., {Grebel}, E.~K., {Binggeli}, B., \& {Glatt}, K. 2007, \apj, 660,
  1186

\bibitem[{{Mastropietro} {et~al.}(2005){Mastropietro}, {Moore}, {Mayer},
  {Debattista}, {Piffaretti}, \& {Stadel}}]{Mast05}
{Mastropietro}, C., {Moore}, B., {Mayer}, L., {et~al.} 2005, \mnras, 364, 607

\bibitem[{{Michielsen} {et~al.}(2008){Michielsen}, {Boselli}, {Conselice},
  {Toloba}, {Whiley}, {Arag{\'o}n-Salamanca}, {Balcells}, {Cardiel}, {Cenarro},
  {Gorgas}, {Peletier}, \& {Vazdekis}}]{Mich08}
{Michielsen}, D., {Boselli}, A., {Conselice}, C.~J., {et~al.} 2008, \mnras,
  385, 1374

\bibitem[{{Moore} {et~al.}(1998){Moore}, {Lake}, \& {Katz}}]{Moore98}
{Moore}, B., {Lake}, G., \& {Katz}, N. 1998, \apj, 495, 139

\bibitem[{{Pedraz} {et~al.}(2002){Pedraz}, {Gorgas}, {Cardiel},
  {S{\'a}nchez-Bl{\'a}zquez}, \& {Guzm{\'a}n}}]{Ped02}
{Pedraz}, S., {Gorgas}, J., {Cardiel}, N., {S{\'a}nchez-Bl{\'a}zquez}, P., \&
  {Guzm{\'a}n}, R. 2002, \mnras, 332, L59

\bibitem[{{Renzini}(2006)}]{Renz06}
{Renzini}, A. 2006, \araa, 44, 141

\bibitem[{{Roediger} \& {Br{\"u}ggen}(2008)}]{Roed08}
{Roediger}, E. \& {Br{\"u}ggen}, M. 2008, \mnras, 388, 465

\bibitem[{{S{\'a}nchez-Bl{\'a}zquez} {et~al.}(2006){S{\'a}nchez-Bl{\'a}zquez},
  {Peletier}, {Jim{\'e}nez-Vicente}, {Cardiel}, {Cenarro},
  {Falc{\'o}n-Barroso}, {Gorgas}, {Selam}, \& {Vazdekis}}]{SB06lib}
{S{\'a}nchez-Bl{\'a}zquez}, P., {Peletier}, R.~F., {Jim{\'e}nez-Vicente}, J.,
  {et~al.} 2006, \mnras, 371, 703

\bibitem[{{Sandage} {et~al.}(1985){Sandage}, {Binggeli}, \& {Tammann}}]{Sand85}
{Sandage}, A., {Binggeli}, B., \& {Tammann}, G.~A. 1985, \aj, 90, 1759

\bibitem[{{Sargent} \& {Turner}(1977)}]{Sarg77}
{Sargent}, W.~L.~W. \& {Turner}, E.~L. 1977, \apjl, 212, L3

\bibitem[{{Silich} \& {Tenorio-Tagle}(2001)}]{ST01}
{Silich}, S. \& {Tenorio-Tagle}, G. 2001, \apj, 552, 91

\bibitem[{{van Zee} {et~al.}(2004){van Zee}, {Skillman}, \& {Haynes}}]{VZ04}
{van Zee}, L., {Skillman}, E.~D., \& {Haynes}, M.~P. 2004, \aj, 128, 121

\bibitem[{{White} \& {Frenk}(1991)}]{WF91}
{White}, S.~D.~M. \& {Frenk}, C.~S. 1991, \apj, 379, 52

\bibitem[{{White} \& {Rees}(1978)}]{WR78}
{White}, S.~D.~M. \& {Rees}, M.~J. 1978, \mnras, 183, 341

\bibitem[{{Yoshii} \& {Arimoto}(1987)}]{YosAri87}
{Yoshii}, Y. \& {Arimoto}, N. 1987, \aap, 188, 13

\end{thebibliography}

\end{document}